\documentclass{vgtc}                          

\ifpdf
  \pdfoutput=1\relax                   
  \usepackage{graphicx}                
  \DeclareGraphicsExtensions{.pdf,.png,.jpg,.jpeg} 
\else
  \usepackage{graphicx}                
  \DeclareGraphicsExtensions{.eps}     
\fi%

\graphicspath{{figures/}{pictures/}{images/}{./}} 

\usepackage{microtype}                 
\PassOptionsToPackage{warn}{textcomp}  
\usepackage{textcomp}                  
\usepackage{mathptmx}                  
\usepackage{times}                     
\usepackage{cite}                      
\usepackage{tabu}                      
\usepackage{booktabs}                  

\newcommand{\modified}[1] 
{\textcolor{black}{#1}}

\vgtccategory{Research}

\vgtcinsertpkg

\title{Folding Rays: a Bimanual Occluded Target Interaction Technique}

\author{DongHoon Kim \thanks{e-mail: donghoon5793@gmail.com}\\ %
        \scriptsize Utah State University %
\and Preston Bruner \thanks{e-mail: preston.bruner1@gmail.com}\\ %
        \scriptsize Utah State University %
\and Isaac Cho \thanks{e-mail: isaac.cho@usu.edu}\\ %
        \scriptsize Utah State University }

\abstract{As Virtual Reality becomes commonplace in the world, it is important for developers to focus on user interaction with the virtual world.
Currently, there are limitations to some selection and navigation techniques that have not yet been completely overcome. Focusing specifically on enhancing ray-casting, we present the advanced technique of folding rays which allows for the selection of occluded targets without any unnecessary physical navigation around a virtual environment. By improving upon current approaches, our technique allows for the selection of these targets without any manipulation of the virtual environment itself using rays that can bend at user-determined points. With their potential to be used in conjunction with teleportation as a virtual navigation technique, folding rays can be used in a variety of scenarios to enhance a user's interactive experience in virtual environments.     
} 

\CCScatlist{
    \CCScatTwelve{Human-centered computing}{Human computer interaction (HCI)}{Interaction techniques}{Pointing}
    \CCScatTwelve{Human-centered computing}{Human computer interaction (HCI)}{Interaction paradigms}{Virtual reality}
}

\begin{document}
\maketitle

\section{Introduction}

Selection is an essential interaction technique in immersive virtual environments. Various techniques have been introduced to translate real-world actions into virtual ones, allowing the user to select virtual objects using familiar paradigms including simple virtual hand \cite{virtual-hand-metaphor} and ray-casting \cite{raycasting-in-vr}. Although these techniques are intuitive and widely employed, their effectiveness is limited when it comes to selecting virtual objects that are not directly visible to the user. When a target object is occluded by other objects, the user is typically required to navigate in virtual environments to bring the occluded object into view before selecting it using one of the selection techniques. In particular, in restrictive real-world scenarios, such as when the user is in a small physical room or sitting in a chair while using a Virtual Reality (VR) headset, physically moving to navigate around the virtual environment can be challenging. To address this limitation, it would be beneficial if the user could select an occluded object without the need for navigation. 

This paper introduces folding rays, a technique that allows the user to select an occluded object. The user can create a ray and then fold it at a desired point. A camera viewport will appear at the fold point, allowing the user to see the ray beyond the fold point. The user may repeat the folding process from within that viewport, giving them full control over the number of folds and the direction of the ray at each fold point. This seamless transition between the views at each fold allows the user to remain comfortably stationary while directing the ray to their liking. This complete control over the folded ray results in a previously unreachable and occluded object becoming accessible to the user without navigation.

\section{Related work}

Ray-casting is a widely used interaction technique, which is a part of ``virtual pointing'', in immersive environments. Typically, rays originate from a virtual representation of an input device so that the user may determine the origin and direction of the ray being cast \cite{raycasting-in-vr}. This gives a level of control when rays are used to select objects further than an arm's reach away from the user. However, the selection of targets with traditional ray-casting is limited to targets that are visible to the user from where they currently are located. 

Several approaches have been developed and implemented to overcome this limitation \cite{9207831}. For example, Alpha Cursor is one such technique that attaches a movable cursor to a selection ray. The user pushes the joystick backward or forward to move the cursor closer or further along the ray. If the cursor goes past an object, that object then becomes transparent, providing a view to the objects behind it \cite{baloup2019raycursor}. Another example called Smash Probe involves spreading out objects that collide with the selection ray to give a view into the world behind them \cite{9207831}. Though this technique allows users to look behind occluding objects, the selection of a desired target may be difficult because of the spreading of objects. Our folding ray-casting technique aims to provide similar insights into a virtual world behind occluding objects but without the need to alter any existing object's properties such as transparency or position. Users will be able to fold their ray around occluding objects without needing to directly interact with them. This will give users an unhindered understanding of the virtual environment around them as no objects will be forcefully manipulated. 

The Heuristic Ray technique introduces a curve target indicator, which looks like a curved ray, to indicate a selected target based on a calculated score\cite{9089485}. While this technique has the potential to select occluded targets, it may face challenges when multiple occluded targets are close together, as the scoring algorithm may make precise selection difficult. We designed our folding ray-casting technique with precision in mind. Users will have full control over the ``bend'' or ``folding'' of their ray, so they will not ever have to question which target they will be selecting when they pull the trigger.

\section{Folding Rays: Implementation}

\begin{figure}[t]
\centering
\includegraphics[width=\columnwidth]{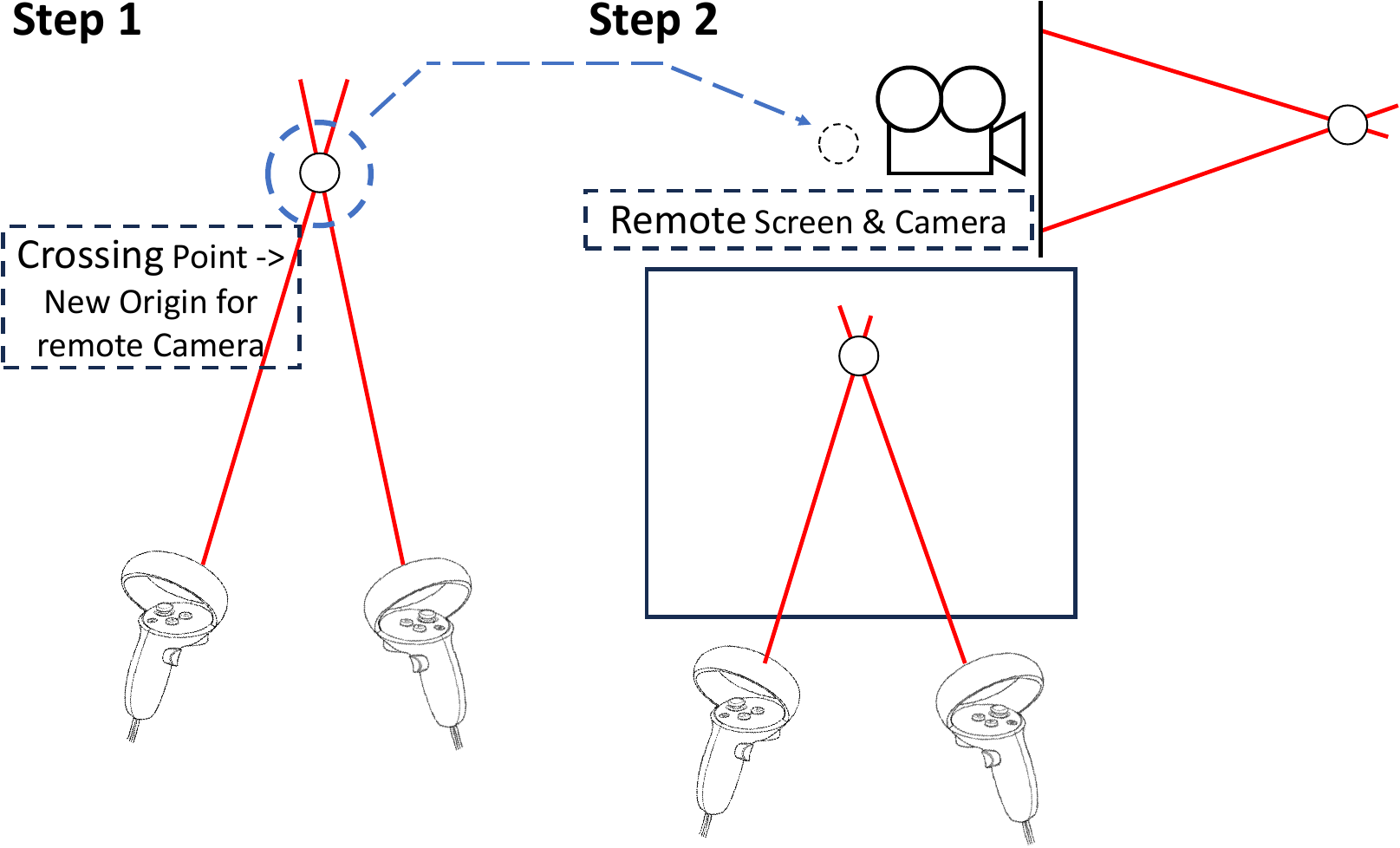}
\caption{
Technique Procedure.
Step 1: Crossing the two rays from  the left and right hands' controllers makes a crossing point. This crossing point allows users to select a specific point in 3D space. 
Pressing the trigger on the dominant hand's controller makes 
a folding point. 
Step 2: After pressing the button, a square screen appears in front of the user that shows the environment from the folding point.
Users can use rays through the screen to interact with objects and make new folding points.
}
\label{figure:implementation}
\end{figure}


Our folding ray-casting technique builds upon the general ray-casting and magic mirror techniques \cite{han2022portal}. Like the general ray-casting approach, our approach utilizes a ray that extends infinitely in one direction, originating from the user's dominant hand (the main ray). The user can select an object that intersects with the ray by pressing a trigger button. 

\begin{figure}[t]
\centering
\includegraphics[width=\columnwidth]{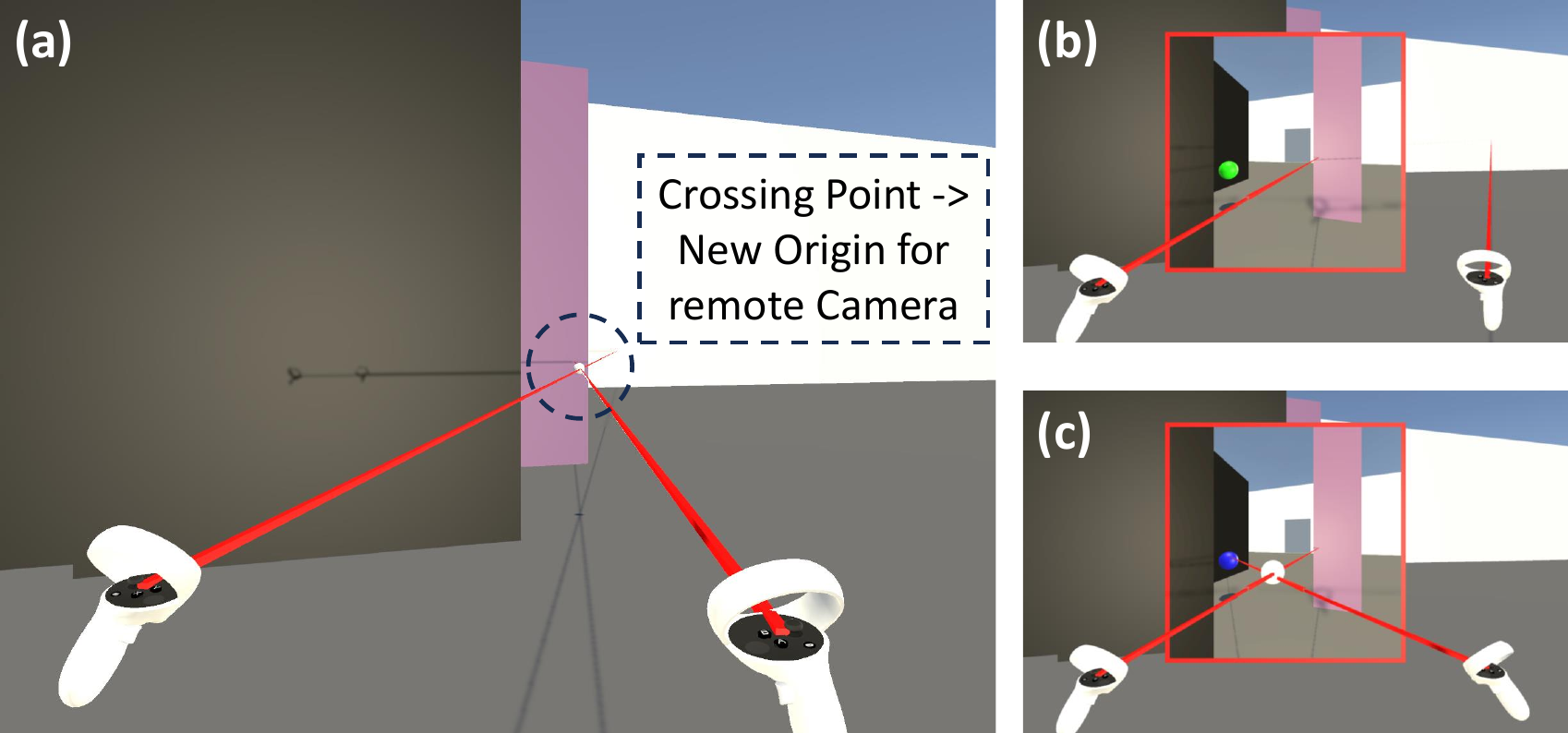}
\caption{
Demo Scene.
(a) The user can not see an occluded target object, which is a green sphere in (b), thus making a crossing point near the wall. (b) The user looks through the screen which appears at the folding point and can now see the target object. (c) The user can select the target now that it is within view.
}
\label{figure:demo}
\end{figure}

Figure \ref{figure:implementation} provides a visual explanation of the folding technique's operation. In our folding ray-casting technique, we include an additional ray originating from the less-dominant hand (the secondary ray). This ray is guided in a similar manner to the main ray by moving the less-dominant hand. The secondary ray is utilized to create a folding point along with the main ray. 
As the secondary ray interests with the main ray within a small distance threshold, a crossing point sphere will appear on the main ray which indicates the folding point. 
Pressing the primary button on the dominant hand's controller creates a camera in front of the point indicated by the sphere, aligning with the user's view direction.
Once the camera is created at the folding point, a camera viewport window \modified{(square screen in Figure \ref{figure:implementation})} appears in front of the user at a comfortable viewing distance of 1.5m. This window shows the view from the folding point. The window follows the user's view direction to allow the user to see a 360 view of the scene from the folding point. The user can create multiple folding points by placing the rays inside the window, and each new folding point updates the window to show the camera at the respective folding point. Figure \ref{figure:demo} shows an example of selecting an occluded target.
Once the target object is visible in the window, the user can point to the main ray to select the object. 


\section{Discussion}
Our folding ray-casting technique has several benefits. 
First of all, it allows for selecting occluded targets without requiring navigation. By allowing users to create multiple folding points along the ray, they can select targets occluded by complex mazes of objects. \modified{This could not be possible with the current interaction techniques which only interact with a non-occluded target \cite{argelaguet2013survey}.
With our technique, users can search where the unknown target object is located and interact with it.
}
This makes the technique very flexible in its use cases. Users with limited space or range of motion will still be able to take full advantage of this technique as no navigation techniques are needed.



Second, the technique addresses the disorientation issue. By implementing the viewport screen, we prevent the user from becoming disoriented while manipulating the camera during the folding process. Instead of taking up the user's entire view with the new camera's position, which has been shown to cause disorientation in applications like teleportation \cite{bowman1997travel}, the viewport screen only takes up a portion of the user's view. Outside of that screen, the user can still see the environment from the original location so that they can easily perceive their surroundings and not lose track of their orientation.

Lastly, the technique can serve as a navigation method. Some current navigation techniques use a form of ray-casting, such as the arc method, to indicate a desired location for teleportation. According to a study that compared multiple teleportation techniques, the arc method was one of the most preferred and most efficient methods for quick navigation\cite{teleportation-methods}. 
Unlike the arc method which limits a single teleportation to a target destination within the user's view, our technique determining the destination involves rays that can fold multiple times, allowing the user to navigate complex environments while teleporting much less frequently. 

There are some limitations with folding rays that could be further explored in the future. 
One limitation is the challenge of precisely selecting a folding point that is located far along the ray, which is a similar limitation in target selection with traditional ray-casting. Additionally, the square viewport screen provides a limited field of view into the environment from the folding point.

\section{Conclusion}

Interacting with occluded objects in a 3D environment is a significant challenge in an immersive environment.
The conventional ray-casting technique is limited in that it can only interact with visible objects from where the user is currently located. Our folding ray-casting technique provides a solution for that limitation using the folding ray and viewport screen.
Using the folding ray technique, users can easily interact with any object regardless of where they are and how occluded the target may be. With their potential uses for both interaction and navigation, folding rays can be used in a wide variety of applications to assist users with limited physical space and range of motion.


\bibliographystyle{abbrv-doi}

\bibliography{final}
\end{document}